%% file: IPRD2019_LAROSA.tex
\documentclass[a4paper,11pt]{article}
\pdfoutput=1 

\usepackage{jinstpub} 

\usepackage{gensymb}

\title{\boldmath The CMS Outer Tracker for the High Luminosity LHC upgrade}

\author[a,1]{A. La Rosa,\note{Corresponding author.}}

\affiliation[a]{CERN EP-DT,\\ CH-1211, Geneve 23, Switzerland}

\emailAdd{alessandro.larosa@cern.ch}

\abstract{The High Luminosity Large Hadron Collider (HL-LHC) at CERN is expected to collide protons at a centre-of-mass energy of 14\,TeV and to reach the unprecedented peak instantaneous luminosity of 5\,$-$\,7.5\,x\,$10^{34}$\,cm$^{-2}$s$^{-1}$ with an average number of pileup events of 140\,$-$\,200. This will allow the ATLAS and CMS experiments to collect integrated luminosities up to 3000\,$-$\,4000\,fb$^{-1}$  during the project lifetime. To cope with this extreme scenario the CMS detector will be substantially upgraded before starting the HL-LHC, with a plan known as CMS Phase-2 upgrade. 
The CMS Tracker detector will have to be replaced in order to fully exploit the delivered luminosity and cope with the demanding operating conditions. The new detector will provide robust tracking as well as input for the first level trigger. This paper is focused on the replacement of the CMS Outer Tracker system, describing the new layout and the technological choices together with some highlights of module assembly and quality assurance aspects.}


\collaboration[c]{on behalf of CMS Tracker Collaboration}

\proceeding{15$^{\text{th}}$ Topical Seminar on Innovative Particle and Radiation Detectors\\
13-17 October 2019\\
Siena, Italy}

\begin{document}
\maketitle
\flushbottom

\input{1_Intro}
\input{2_OT_for_HL-LHC}
\input{3_Modules}
\input{4_Assembly_QA}
\input{5_Conclusion}

\newpage

\end{document}

%% file: 1_Intro.tex
\section{Introduction}
\begin{figure}[h!]
\centering
\includegraphics[scale=0.3]{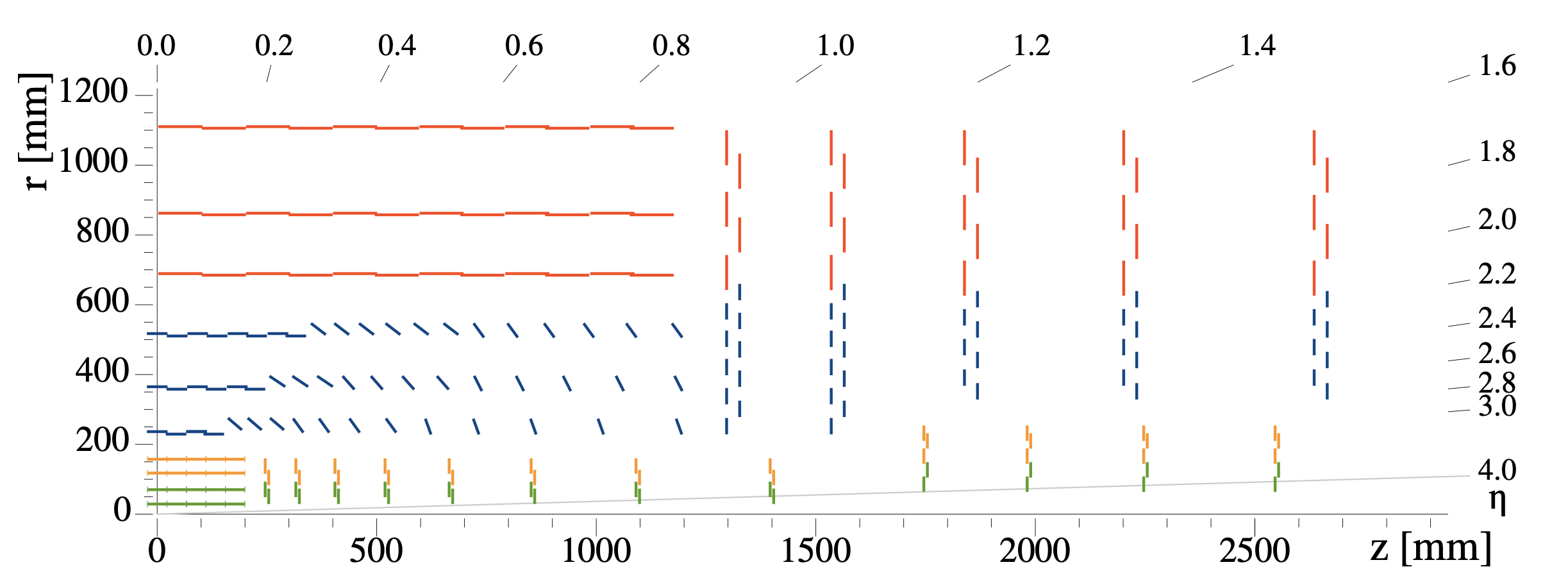}
\caption{Sketch of one quarter of the tracker layout in r-z view. In the Inner Tracker the green lines correspond to pixel modules made of two readout chips and the yellow lines to pixel modules with four readout chips. In the Outer Tracker the blue and red lines represent the two types of modules, respectively equipped with pixel-strip and strip-strip sensors ~\cite{TKTDR}.}
\label{fig:TK-layout}
\end{figure}

With the upgrade of the LHC instantaneous luminosity (HL-LHC~\cite{HL-LHC}) up to 7.5\,x\,$10^{34}$\,cm$^{-2}$s$^{-1}$, ATLAS~\cite{ATLAS} and CMS~\cite{CMS} will need to operate at an average up to 200 interactions per 25\,ns beam crossing time and expect to obtain up to 4000\,fb$^{-1}$ of integrated luminosity. To achieve their physics goals the experiments will need to improve the tracking resolution and the ability to selectively trigger on specific physics events at reasonable thresholds.\\ 
The currently installed CMS tracker detector (strip~\cite{CMS} and pixel~\cite{Pixel} systems) will not withstand the expected radiation dose and their performance will be inadequate to cope with the demanding operating conditions at HL-LHC. For these reasons, CMS has foreseen to replace the tracking system during the LHC Long Shutdown 3 (2023-26). The new CMS tracker detector~\cite{TKTDR} will consist of about 220\,m$^{2}$  of silicon detector and will be composed of two sub-detectors: the Outer Tracker (OT) and the Inner Tracker (IT).  The OT detector will consist of  a combination of silicon double-sided (strip-strip and pixel-strip) modules, while the IT detector will be made by silicon pixel modules. A sketch of one quarter of the new CMS tracker layout in r-z view is shown in Figure\,\ref{fig:TK-layout}.

%% file: 2_OT_for_HL-LHC.tex
\section{The CMS Outer Tracker for HL-LHC}
\begin{figure}[h!]
\centering
\includegraphics[scale=0.45]{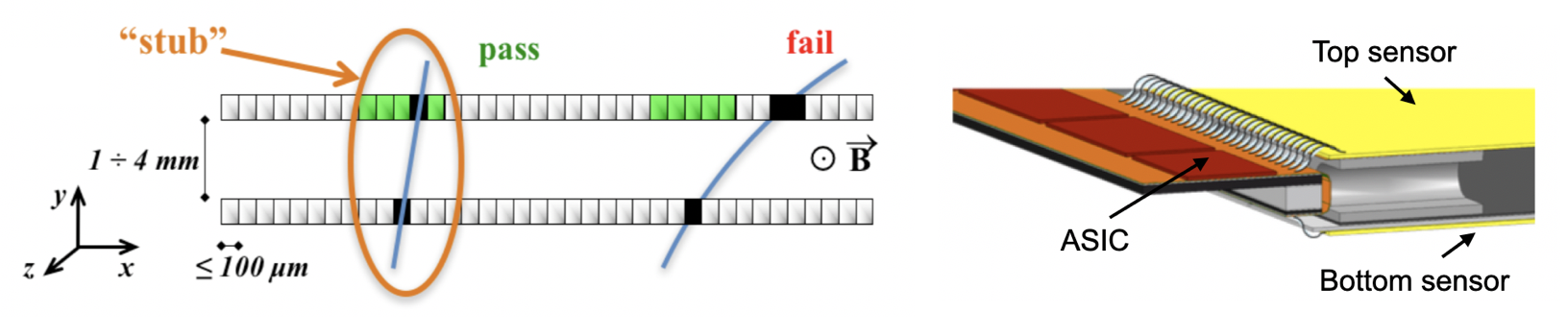}
\caption{Sketch of the track stub finding principle (left) and the  p$_{\mathrm{T}}$ module concept (right). A track passes both sensors of a module. A low momentum track falls outside the acceptance window and produces no stub~\cite{TKTDR}.}
\label{fig:pT-concept}
\end{figure}

\begin{figure}[h!]
\centering
\includegraphics[scale=0.3]{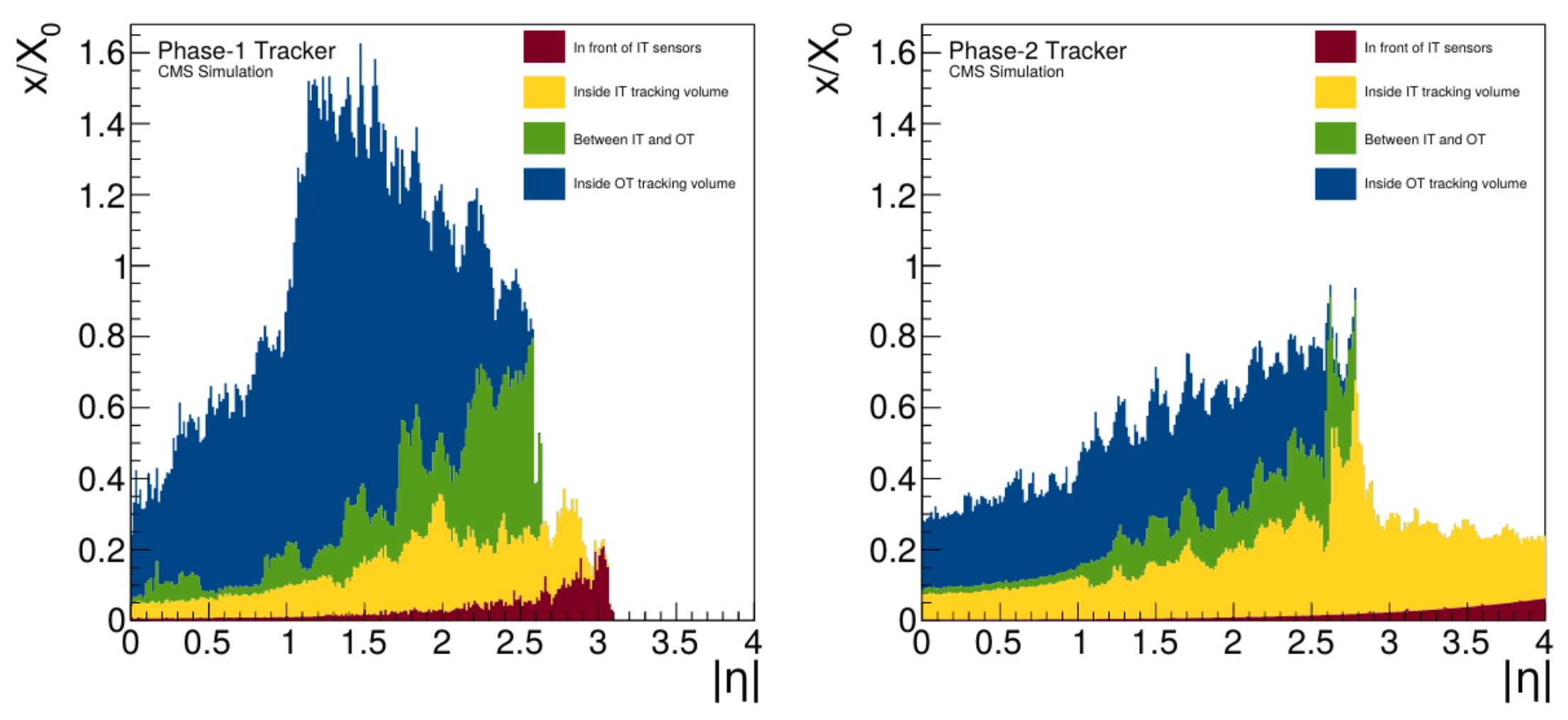}
\caption{Material budget for the currently installed CMS tracker (left) and the new tracker (IT and OT detectors) (right)~\cite{TKTDR}.}
\label{fig:material-budget}
\end{figure}

\begin{figure}[h!]
\centering
\includegraphics[scale=0.3]{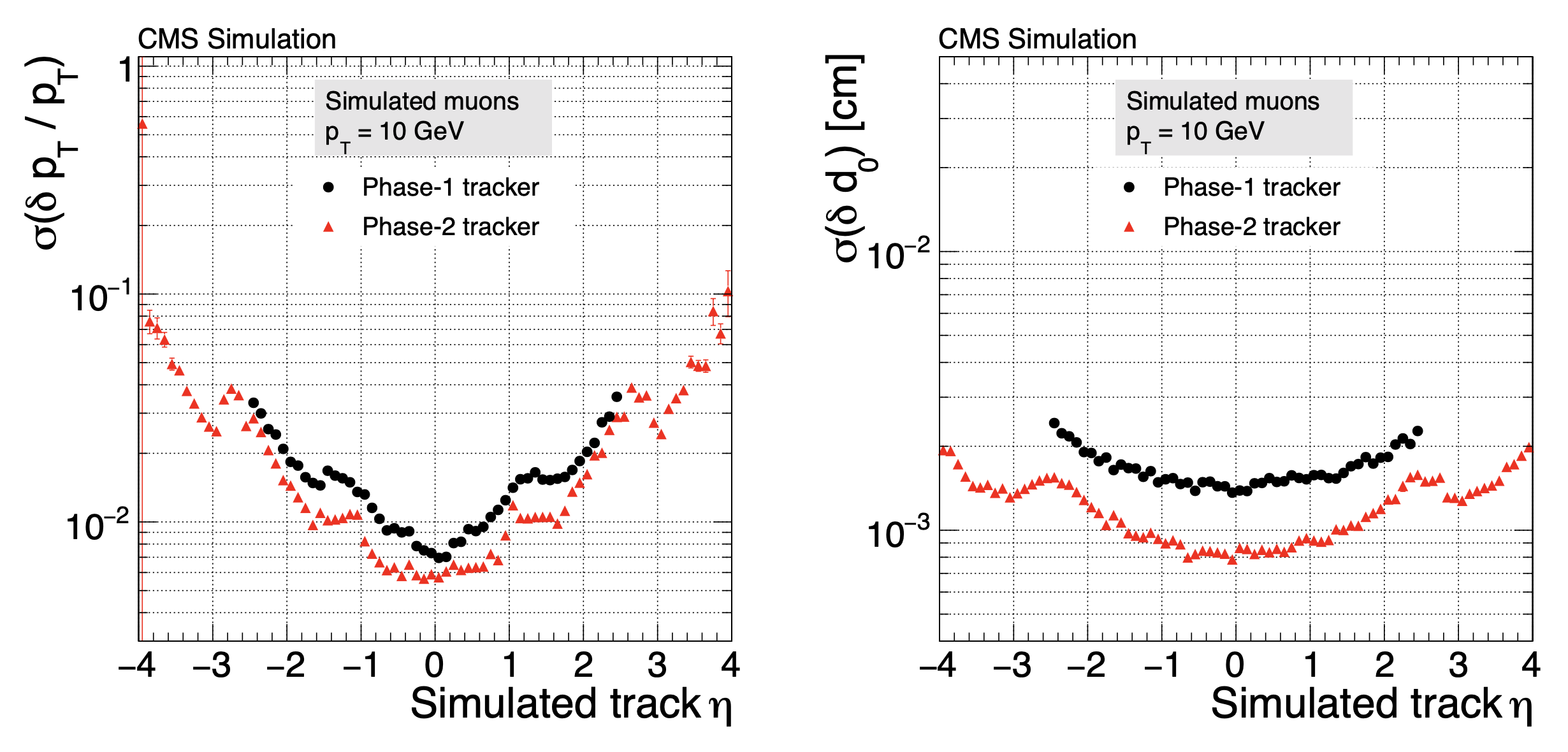}
\caption{Relative resolution of the transverse momentum (left) and resolution of the transverse impact parameter (right) as a function of the pseudorapidity for the Phase-1 (black dots) and the new  Phase-2 tracker (red triangles), using single isolated muons with a transverse momentum of 10\,GeV ~\cite{TKTDR}.}
\label{fig:performances}
\end{figure}

The Outer Tracker detector for the HL-LHC consists of six barrel layers and five endcap disks per side, and can be subdivided in four main large structures: the TB2S (Tracker Barrel with 2S modules), TBPS (Tracker Barrel with PS modules) and two TEDDs (Tracker Endcap Double Disks). The basic unit of the detector is the so called p$_{\mathrm{T}}$ module. Two type of p$_{\mathrm{T}}$ modules are employed in the detector: 2S and PS. Both module types have two closely spaced silicon sensors. The 2S modules has two sensors with micro-strips, while the PS modules consist of one sensor with micro-strips and one sensor with macro-pixels. The two sensors are read out by common front-end ASICs capable of correlating the hits from the two sensors. The main feature of the module is its ability to provide tracking information to the first CMS trigger level (L1) by measuring transverse momentum of tracks. The use of tracking information in the L1 trigger implies that the detector has to send out self-selected information at every bunch crossing. The p$_{\mathrm{T}}$ module's functionality relies upon local data reduction in the front-end electronics, in order to limit the volume of data that has to be sent out at 40\,MHz. 
This is achieved by p$_{\mathrm{T}}$ modules capable of rejecting signals from particles with transverse momentum smaller than a given p$_{\mathrm{T}}$ threshold.
As sketched in Figure\,\ref{fig:pT-concept}, thanks to the strong CMS magnetic field, charged particles that go through the detector are bent in the transverse plane by a p$_{\mathrm{T}}$ dependent angle.
The module front end ASIC receives the hits' location, measures the local distance, and compares that to a predefined acceptance window to select candidates with high p$_{\mathrm{T}}$.
A local track segment (so called track stub) is formed and subsequently pushed out to the L1 trigger system at every bunch crossing.  The backend track finder system receives the stub data from the individual detector modules and performs track finding  in two steps, pattern recognition and track fitting\,~\cite{TKTDR}.
\newline
In a standard flat geometry a track may not cross both bottom and top sensor of a module for a given $\eta$ (assuming vertex constrains). To enable a good hermetic coverage and a better trigger performance, a tilted geometry is employed in the tracker barrel equipped with PS modules (see Figure\,\ref{fig:TK-layout}).
Despite the  increased number of readout channels in the new tracker the estimated material budget of the Outer Tracker shows a significant reduction as is reported in Figure\,\ref{fig:material-budget}, where the currently installed tracker (Phase-1) vs.\,the new tracker (Phase-2) comparison is shown. The key features to achieve this material budget reduction are: a reduced number of layers, an optimised routing of the services, a use of light weight material for support structures, a low-mass CO2 cooling and a use of DC-DC converters\,~\cite{TKTDR}.
\newline
From a tracking parameter resolution prospective the new tracker will be also more performant than its predecessor. Figure\,\ref{fig:performances} shows respectively the p$_{\mathrm{T}}$ resolution (left) and impact parameter resolution (right) comparison between the currently installed tracker (Phase-1) vs.\,the new tracker (Phase-2).

%% file: 3_Modules.tex
\section{The p$_{\mathrm{T}}$ modules}
\begin{figure}[h!]
\centering
\includegraphics[scale=0.43]{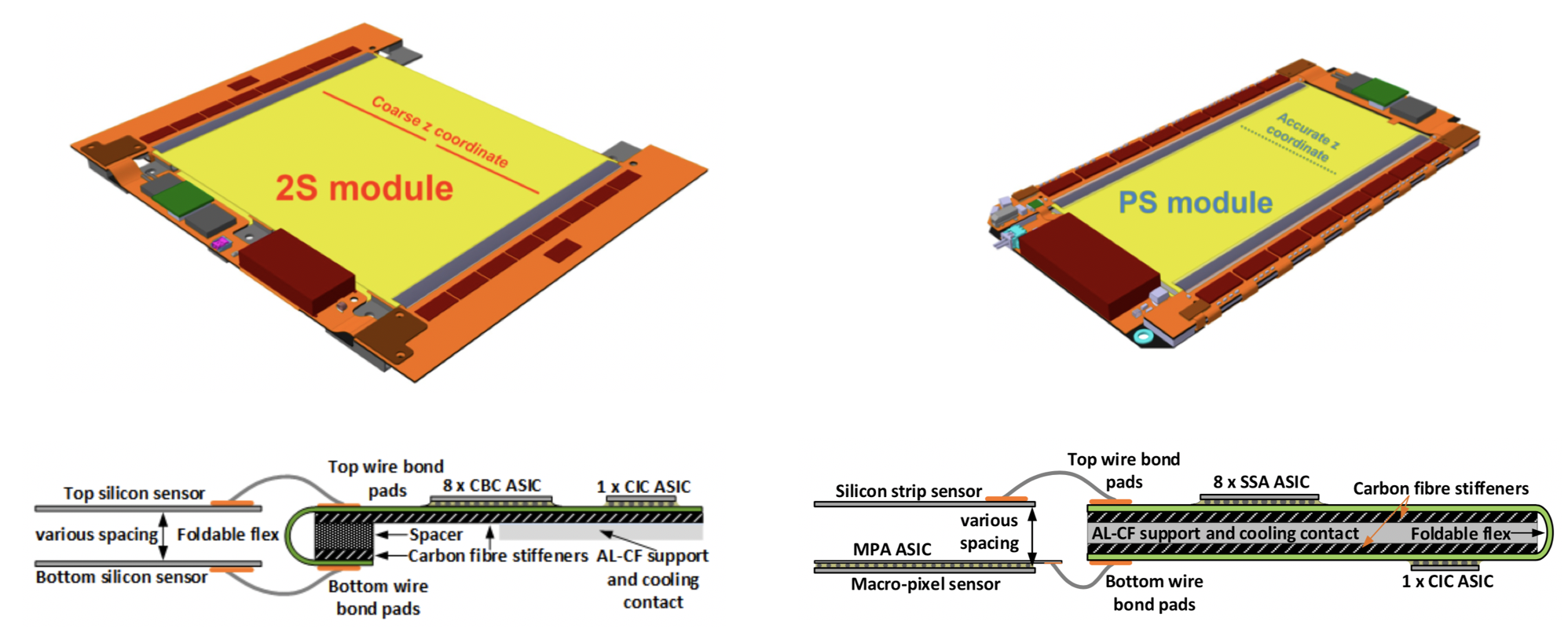}
\caption{The 2S module (left) and PS module (right) of the Outer Tracker. Shown are views of the assembled modules (top), and sketches of the front-end hybrid folded assembly and connectivity (bottom) ~\cite{TKTDR}.}
\label{fig:pT-modules}
\end{figure}

The p$_{\mathrm{T}}$ modules are equipped with on-module electronics and services hybrids, and they are stand-alone units that are directly connected to the backend electronics with no intermediary aggregator system. A sketch of both module types are shown in thes Figure\,\ref{fig:pT-modules}.
\newline
The sensor technology chosen for both sensors of the p$_{\mathrm{T}}$ modules is the Float Zone (FZ) silicon n-in-p with an active sensor thickness of 290\,$\mu$m (referred later as FZ290).  An extensive irradiation and characterization program showed that FZ290 provides sufficient signal-to-noise at the standard operation voltage of 600\,V and the expected maximum fluence after 3000\,fb$^{-1}$. For scenarios up to and beyond 4000\,fb$^{-1}$, an increase to an operation voltage of 800\,V would allow FZ290 to maintain adequate performance even at the most exposed locations.
\subsection{2S module description}
The 2S module consists of two silicon micro-strip sensors. Each sensors has two columns of 1016 strips each with single cell size of 5\,cm\,$\times$\,90\,$\mu$m. The micro-strip sensors are read out by the CMS Binary Chip (CBC) implemented in 130\,nm CMOS technology. Each chip reads 254 strips (127 from bottom and 127 from top sensor strips) and performs correlation of hits between two sensors forming track stubs and sending the stub data out at each bunch crossing. In order to enable stub finding across boundaries the CBC exchanges data with its neighbour chip. As sketched in the left of Figure\,\ref{fig:pT-modules}, eight CBCs are hosted on each front-end hybrid per side. The CBCs send data to the Concentrator Integrated Circuit (CIC), designed in 65\,nm CMOS technology, that performs data sparsification, formats the output data, and sends them to the service hybrid. The CIC processes both the track stub data at 40 MHz and the detector payload which is sent out after a L1 accept signal with a maximum rate of 750\,kHz. As shown in top left of Figure\,\ref{fig:pT-modules}  the modules host the service hybrid, with the LpGBT~\cite{LpGBT}, Versatile Link chips~\cite{VTRx}, DC-DC converters~\cite{DCDC} and HV distribution circuitry. 
The input voltage provided by the power supply system in the experimental cavern will be in the range 10\,-\,12\,V. 
Each module has two DC-DC converters to convert to the voltages used by the various module components. In a first step the voltage is transformed from 10\,-\,12\,V down to 2.5\,V which can be used directly for the biasing of the optical electronics.
A second DC-DC converter stage transforms the 2.5\,V down to the voltage required by the ASICs (between 1.0\,V and 1.25\,V depending on ASIC).
The data from both (left and right) front-end hybrids are merged and sent via a single optical fibre to the back-end electronic system. To enable a homogeneous p$_{\mathrm{T}}$ (>\,2\,GeV/c) filtering in different detector regions different sensor spacings are chosen. For the 2S module two spacings are selected: 1.8\,mm and 4.0\,mm.
\subsection{PS module description}
The PS module consists of one silicon micro-strip sensor organised in two columns of 960 strips each with single cell size of 2.5\,cm\,$\times$\,100\,$\mu$m, and a macro-pixel sensor with a matrix of 32\,$\times$\,960 pixels with a pixel size of 1.5\,mm\,$\times$\,100\,$\mu$m. The strip sensor is read out by the Strip Sensor ASIC  (SSA) containing 254 channels, while the macro-pixel sensor is readout  by the Macro-Pixel ASIC (MPA); both chips are implemented in 65\,nm CMOS technology. 
A total of 16 MPAs are bump-bonded to the  pixel sensor resulting in an assembly called macro-pixel sub-assembly (MaPSA). 
To build the stubs, the hit correlations of micro-strip and macro-pixel takes place in the MPA chip. The SSA chip, which sits on the front-end hybrid, as the CBC for the 2S module, processes the sensor signals and sends sparsified cluster data to the corresponding MPA chip at each bunch crossing. 
The MPA processes and sparsifies the hits from each macro-pixel. It correlates the bottom macro-pixel sensor hits with the data received from the SSA strips and builds stubs. Connections to and from the front-end hybrids are through wire bonds. As in the 2S module, the CIC on the front-end hybrid buffers and aggregates the stub and cluster data received from the MPAs, and sends them to the service hybrid that in this case due to space constrains is divided in two: readout hybrid and power hybrid as shown in the top right of Figure\,\ref{fig:pT-modules}. 
The transfer scheme and data formats are very similar to those used in the 2S module, allowing the CIC chip to be used for both module types. 
Also for the PS module, in order to have an homogeneous p$_{\mathrm{T}}$ filtering in different detector regions different sensor spacings are chosen: 1.6\,mm, 2.6\,mm and 4.0\,mm.

%% file: 4_Assembly_QA.tex
\section{Module construction and quality assurance plan}
\begin{figure}[h!]
\centering
\includegraphics[scale=0.45]{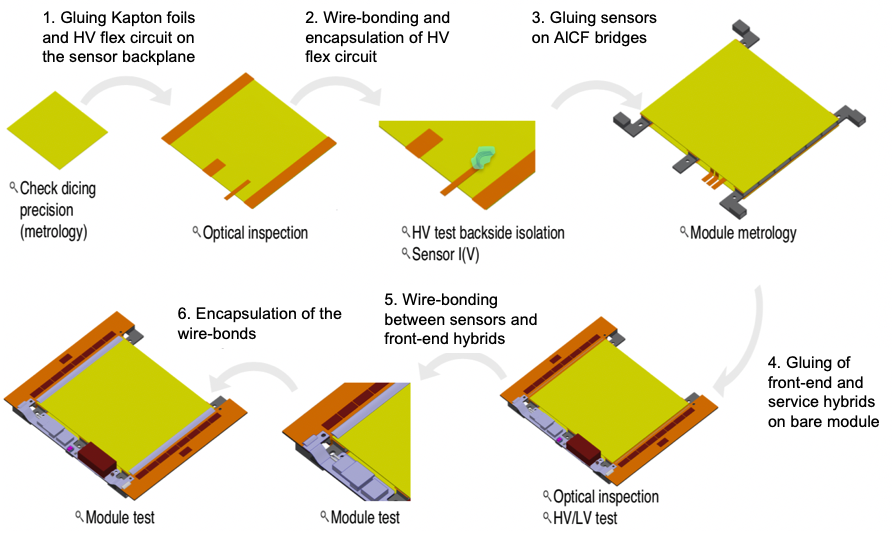}
\caption{Sketch of the 2S module assembly procedure ~\cite{SM}.}
\label{fig:procedure}
\end{figure}

\begin{figure}[h!]
\centering
\includegraphics[scale=0.4]{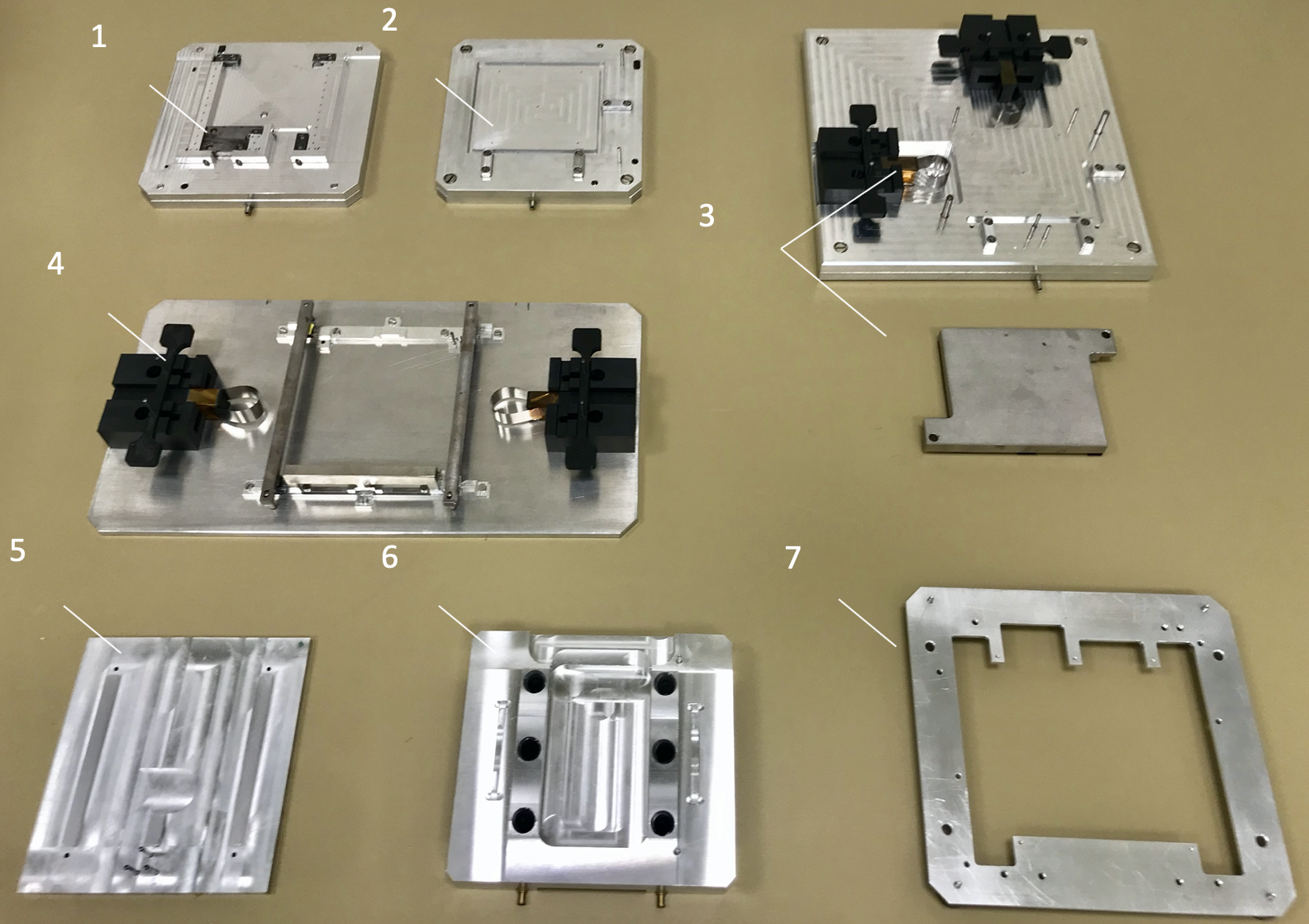}
\caption{Photos of jigs used to assembly 2S modules. The jigs shown are: 1) Polyimide HV positioning-holding jig;
2) Sensor backplane gluing plate;
3) Sensor gluing jig (including weight plate);
4) Readout and service hybrids gluing jig;
5) Glue transfer plate for long and stump bridges;
6) Wire-bonding jig;
7) Module carrier.
 }
\label{fig:jigs}
\end{figure}

\begin{figure}[h!]
\centering
\includegraphics[scale=0.45]{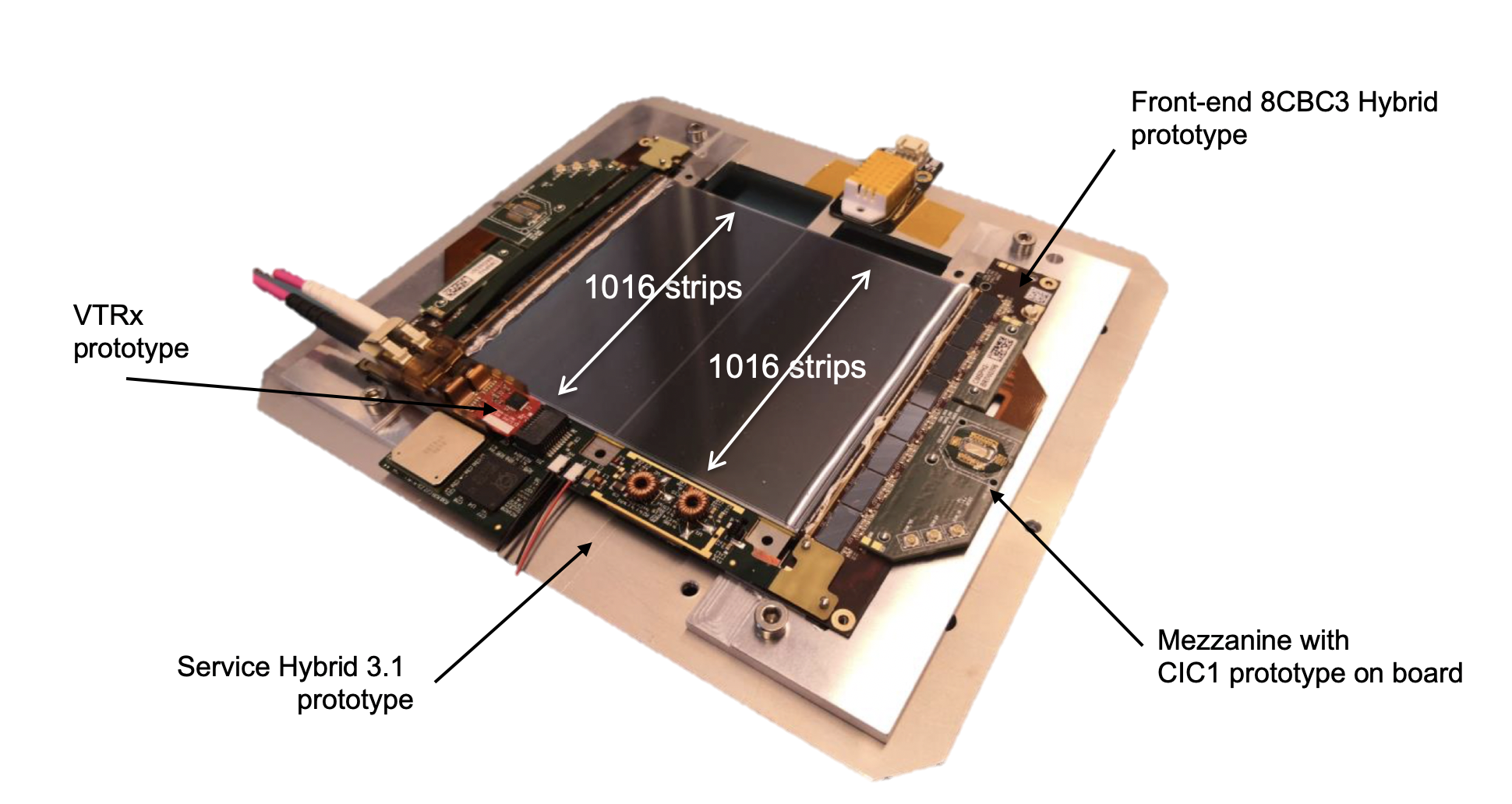}
\caption{Photo of the most recent 2S module prototype.}
\label{fig:2Smodule}
\end{figure}

The module construction consists  of the assembly of the following components: two strip sensors, three spacers (called bridges) of 
an aluminium carbon fibre composite material  (AlCF) having lower mass and better thermal conductivity than aluminium, and three hybrids for the 2S modules; one strip sensor plus one MaPSA, two spacers of AlCF, four hybrids, and a baseplate for the PS modules. As an example in Figure\,\ref{fig:procedure} an overview of the assembly module procedure for the 2S module is shown. 
In this case, the first step of the module assembly procedure consists of gluing Kapton foils and the HV bias flex circuits to the backplane of the sensors.
The most critical step for alignment precision is the sensor gluing onto the AlCF bridges. For this a jig that has three precision machined stops to align the edges of the sensors  (Figure\,\ref{fig:jigs}, jig n.\,4) is used. 
After this step, a measurement of the alignment accuracy is performed. 
The hybrids are then glued to the extremity of the AlCF bridges in another dedicated jig (Figure\,\ref{fig:jigs}, jig n.\,5). 
The gluing of hybrids is followed by the wire bonding with aluminium wire of  25\,$\mu$m diameter, which is performed first on the top side and then on the bottom side. 
After wire bonding a full electrical test of the module is carried out to ensure that all strips are properly connected to the readout chips and that the readout chips and the service hybrid are fully functional. For this test the module is then moved to the module carrier, which is designed to provide protection to the module during the handling needed during testing. If the module passes the test, it stays on the module carrier for the  wire bond encapsulation and the final electrical acceptance test.
The specifications for the sensor to sensor alignment during assembly are: a distance perpendicular to the strips of $\Delta$x\,<\,50\,$\mu$m, a distance along the strips of $\Delta$y\,<\,100\,$\mu$m, and a tilt angle between the strips (strips and macro-pixels) smaller than 400\,$\mu$rad (800\,$\mu$rad) in 2S (PS) modules.
The construction of both 2S and PS modules is based primarily on manual jig-based assembly technique, and Figure\,\ref{fig:jigs} shows an overview of what has been developed and proved to work for building 2S modules.
\newline
The module prototyping has been ongoing for several years. 
For the 2S module the baseline design has been proven to work as several functional modules built in different CMS institutes. Figure\,\ref{fig:2Smodule} shows the last module  built with the most recent prototype components. For the PS module so far only few mechanical modules have been assembled using prototype jigs, but:
the MPA was bonded with macro-pixel sensor and successfully tested in beam, a mini-module containing two SSA chips has been commissioned as well by checking the front-end functionality in lab and beam-test, and also the communication between a single MPA and a single SSA has been successfully accomplished.
\newline
Module QA is a critical component of the production process considering the huge number of identical or similar modules (>13500) that need to be built, their inaccessible location once installed, and the   extreme environmental conditions in which they will operate.
The module QA plan will include the component production, the assembly procedures, and the testing of modules. To monitor and evaluate the production quality the modules will be tested directly after assembly.  All the assembly centres that will build the same kind of modules are required to have a common set of assembly tooling and procedures. A traveller  document system will be used to track the components on the module and the results of checks and tests occurred during the module assembly. The most important assembly data are: dimensional measurements, visual inspection results, and electrical test results of hybrids and modules. These data will be entered into a construction database that will allow a full traceability of the module up to their final integration into the detector.
The module production will be preceded by a pre-production, where the modules will undergo a rigorous testing process in order to check that they meet the specifications of the project. Any components or modules that will fail the quality process will undergo a failure analysis to reveal the root causes of the failure and so allowing modifications to the components, tooling or procedures to solve the problems.

%% file: 5_Conclusion.tex
\section{Conclusion and outlook}
To cope with the higher pile-up and radiation environment foreseen at the HL-LHC the currently installed CMS Tracker detector needs to be fully replaced. 
The new tracker will consist of about 220\,m$^{2}$  of silicon detector and will be made up of  the Outer Tracker using modules containing pairs of closely spaced sensors, and the Inner  Tracker with silicon pixel sensors.\\
The main features of the Outer Tracker are related to the high granularity of the detector to keep channel occupancy low, the radiation hardness to withstand the expected fluence and total ionising dose, the low material budget for improving tracking performance in the high pile-up scenario and last but not least  provide tracking information to the first stage of trigger system by measuring transverse momentum of tracks to reduce data volume.
The core of the Outer Tracker is the so called p$_{\mathrm{T}}$ module that has the capability to identify hits between the two closely spaced silicon sensors. The module is a stand-alone unit  equipped with all needed read out and service electronics fully embedded  on it.  
The module prototyping has been progressing for many years resulting in the assembly of functional modules. Many modules have been produced in several different CMS institutes so far with successful test results that have proven the baseline module design and assembly procedure.\\
In conclusion, the CMS Outer Tracker project is actually well on-track and is ramping-up towards the pre-production.

%% file: IPRD2019_LAROSA.bbl
\begin{thebibliography}{99}
\bibitem{HL-LHC} G.Apollinari et al., High-Luminosity Large Hadron Collider (HL-LHC), CERN Yellow Rep. Monogr. 4 (2017) 1-516.
\bibitem{ATLAS} ATLAS Collaboration, The ATLAS Experiment at the CERN Large Hadron Collider, JINST 3 (2008) S08003.
\bibitem{CMS} CMS Collaboration, The CMS Experiment at the CERN LHC, JINST 3 (2008) S08004.
\bibitem{Pixel} CMS Collaboration, CMS Technical Design Report for the Pixel Detector Upgrade, CMS Technical Design Report CERN-LHCC-2012-016, CMS-TDR-11, 2012.
\bibitem{TKTDR}CMS Collaboration, The Phase-2 Upgrade of the CMS Tracker - Technical Design Report, CERN-LHCC-2017-009, CMS-TDR-014.
\bibitem{LpGBT} CERN, LpGBT specification document, https://espace.cern.ch/GBT-Project/LpGBT/Specifications/LpGbtxSpecifications.pdf.
\bibitem{VTRx} J. Troska, et al., The VTRx+, an Optical Link Module for Data Transmission at HL-LHC, PoS (TWEPP-17)048.
\bibitem{DCDC} CERN, Development of DC-DC converter at CERN,  http://project-dcdc.web.cern.ch/project-dcdc/Default.html.
\bibitem{SM} S. Maier et al., Module Prototyping for the Phase II Upgrade of the CMS Outer Tracker, presented at the Forum on Tracking Detector Mechanics in 2018 [https://indico.cern.ch/event/695767].




\end{thebibliography}
